\documentclass{aa}
\usepackage{graphics}
\begin{document}
\def\cd{cd$^{-1}$}
\def\cds{cd$^{-1}$\,}
\def\kms{km~s$^{-1}$}

   \thesaurus{06     
              (03.13.2,
               03.20.8,
               03.20.4,
               08.09.2 X Cae,
               08.15.1,
               08.22.2)} 
   \title{
Simultaneous intensive photometry and high resolution spectroscopy of $\delta$ 
Scuti stars IV. An improved picture of the pulsational behaviour of X Caeli.
\thanks{Based on observations collected at European Southern Observatory 
(Proposal 58.E-0278)}}
   \author{L. Mantegazza
          \inst{1}
          \and
          F.M. Zerbi\inst{1}
          \and
          A. Sacchi\inst{2}
          }
   \offprints{L. Mantegazza}

   \institute{Osservatorio Astronomico di Brera
              Via E. Bianchi, 46 I-23807 Merate Italy\\
              \thanks{e-mail: luciano@merate.mi.astro.it,zerbi@merate.mi.astro.it}
          \and
              Universit\'a di Pavia, Dipartimento di Fisica Nucleare e
              Teorica. Via Bassi 6, I-27100 Pavia Italy\\
              \thanks{e-mail: sacchi@merate.mi.astro.it}
              }

   \date{}
   \maketitle
\markboth{L. Mantegazza et al.: Pulsational behaviour of X  Caeli}{}
   \begin{abstract}
The $\delta$ Scuti star X Caeli has been the target of a simultaneous
photometric (14 consecutive nights) and spectroscopic campaign (8 consecutive
nigths). From the analysis of light curves we were able to pick up 17
frequency components, most of which were already detected in two previous
campaigns. The comparison with the results of the previous campaigns shows that
while some terms are rather stable (in particular the dominant mode at
7.39 \cds) others have conspicuous amplitude variations. 14 photometric terms
have been also detected in the radial velocity curve or in the analysis
of the line profile variations. There are no spectroscopic terms
without photometric counterparts, and this means that there are no
high--degree modes as observed in other $\delta$ Scuti stars.
The simultaneous fit of light and line profile variations has allowed
the estimation of the inclination of rotational axis (about $70^o$) and 
the
$\ell,m$ identification for many modes. In particular there is
clear evidence that the two shortest period modes are retrograde.
Rather reliable
results were found for the dominant mode which  has $\ell=1$, $m=-1$.
The resulting physical parameters of its pulsation are in good agreement
with the prediction of theoretical models and suggest for this star
a mixing length parameter of about 0.5.
Finally the fundamental stellar physical parameters and their
refinement are discussed in the light of the identification of the 7.47 \cds~
term as a radial mode.


\end{abstract}

%

\section{Introduction}

X Cae has been recently the target of two campaigns: the first
photometric in 1989 (Mantegazza \& Poretti, 1992, Paper I) and the second
both photometric and spectroscopic in 1992 (Mantegazza \& Poretti 1996,
Paper II).

From the photometric data obtained in these campaigns it has been
possible to single out 14 terms
due to at least 6 independent pulsation modes and to some harmonics
and non--linear couplings among the strongest ones.
With the additional information supplied by the high resolution spectroscopy,
an attempt has been performed to identify at least some of these modes.
However, due to the rather short baseline of spectroscopic data (4 consecutive
nights), only very preliminary results have been obtained. The only really
confident result was that the dominant mode (7.39 \cds) has $m=-1$
and $\ell=$ 1 or 2.
A successive improved analysis of these data (Mantegazza \& Poretti, 1998)
 favoured the
choice of $\ell=1$ and $m=-1$, and suggested that
the rotational axis inclination is larger than $45^o$.
About the other modes the conclusion was that they
are probably non-radial with $\ell=2\pm1$, while high--degree sectorial modes,
which seem to be rather common in fast
rotating $\delta$ Scuti stars (Mantegazza, 1997; Kennelly et al., 1998), 
were not observed.
The comparison  between the photometric data of the two campaigns shows that 
while
the dominant mode has a very stable amplitude, those of other modes show
a certain degree of variability.
The line profiles showed the presence of a narrow absorption core,
which was attributed to the presence of a circumstellar shell,
even if the possibility of a companion could not be completely ruled out.

In this paper we shall present and discuss the results of
a new simultaneous photometric--spectroscopic campaign performed in 1996
in which the spectroscopic baseline has been doubled with respect to the
previous one (8 consecutive nights).  The longer spectroscopic baseline
and the simultaneous collection of photometry and spectroscopy allowed us
to perform the analysis through a synergic approach, first formalized in Bossi 
et al (1998): such an approach seems the most rewarding in the analysis
of non-radial multi-mode pulsators.

\section{Physical parameters}

The stellar physical parameters were derived in Paper II using
$uvby\beta$ photometric indices. However the stellar luminosity, as
estimated from the Hipparcos parallax  ($M_{bol}=1.30\pm0.15$),
does not agree with that derived from this approach ($M_{bol}=0.73\pm0.3$), 
therefore we decided to reconsider the estimation of all these parameters.

Moon \& Dworetsky (1985) calibration applied to the $uvby\beta$ data
extracted from the catalogue by Hauck \& Mermilliod (1990)
supplies $T_{eff}=6900^oK$ and $\log g=3.50$ (this includes also the correction
for metallicity effects, Dworetsky \& Moon , 1986);
at the same time the very recent calibration
of Geneva photometric indices by Kunzli et al. (1997) applied to the data
by Rufener (1988) gives $T_{eff}=6860\pm60^oK$ and $\log g=3.76\pm0.14$.
We see that, while we can assume $T_{eff}=6900\pm100^oK$, there is a
certain disagreement on $\log g$.
From the Hipparcos luminosity and the above derived photometric temperature 
we can estimate 
a radius $R=3.43\pm0.32 R_\odot$, and we can also see that, according 
to the theoretical
models by Shaller et al. (1992), these two quantities fit a stellar mass
of $M=2.0\pm0.1$.
Hence we derive $\log g=3.66\pm0.09$, which is just in between
the two photometric estimates, and therefore in the following we shall adopt
this value.
With these physical parameters we can estimate that the frequency of the
fundamental radial mode should be $\nu_{F}=6.7\pm0.7$\cds.

\section{Observations and data processing}

\subsection{Photometry}

Photometry of the star X Caeli was performed with the ESO 50 cm telescope at
la Silla, Chile, for 14 consecutive nights (November 15--28, 1996), 
covering 100 hours
of useful observation time. HR 1766 and HR 1651 where chosen as comparison
stars as their stability was already confirmed by the previous campaigns
(Mantegazza and Poretti, 1996). 

Although air transparency in la Silla was rather stable during the observing
nights, the data were processed using the instantaneous extinction coefficients
technique developed by Poretti and Zerbi (1993) in order to deal with the 
residual
intra-night transparency variations. The reduced photometric data consist
of 916 and 889 differential measurements in Johnson's $V$ and $B$ bands
respectively.

The analysis of the magnitude differences between the two comparison stars
 confirmed their stability and showed that the photometric quality of
the nights was acceptable. The standard deviation of the $B$ and $V$ 
differential
time series is 3.67 and 3.57 mmag respectively and the frequency analysis of
these series by means of the least square technique (Vani\^cek, 1971) 
showed a flat spectrum:
the highest peaks show amplitudes not exceeding 1.4 \% and 2.2\% (0.32 and
0.37 mmag semi--amplitude in the corresponding sinusoidal signal) respectively.

\subsection{Spectroscopy}

The spectroscopic observations have been performed in the Remote Control
Mode at the La Silla Observatory  (ESO) using the Coud\`e
Echelle Spectrograph (CES) attached to the Coud\`e Auxiliary Telescope
(CAT) during 8 consecutive nights (November 21-28, 1996).
The spectrograms, which cover the 4483--4533 \AA~ region,
were acquired by means  of the ESO CCD \#38. The resulting
resolution was of 0.018 \AA/pixel and the effective resolving
power, as measured from spectrograms, was 55000.

225 spectrograms of X Cae with
exposure times of 15 minutes have been gathered in total. They cover about
61 hours of observing time.
The data reduction was performed using the MIDAS package developed at ESO.
The spectrograms have been normalized by means of internal quartz lamp
flat fields, and calibrated into wavelengths by means of a Thorium lamp.
The normalization at the continuum has been performed by fitting a third
degree polynomial to the continuum windows in  the region containing
the three adjacent spectral
lines which already have been used in the previous study (i.e. the 
normalization was limited to the 4497--4517 \AA~region ).
Finally the spectra have been shifted and rebinned in order to
remove the observer's motion.

The mean standard deviation of the pixel values in the continuum regions allows
the estimation of the signal--to--noise ratio of the spectrograms. 
The resulting
average value at the continuum level is about 230.

Fig. 1 shows in the upper panel the average spectrum. We can notice that
the spectral lines have at the center the narrow absorption core already
present in the spectrograms of the previous campaign.
The lower panel shows the standard deviation of the individual spectrograms
about the average one. The figure clearly shows the variability of the line 
profiles; we can also notice the close similarity with Fig. 4 of 
paper II.
\begin{figure}[]
\resizebox{9cm}{!}{\includegraphics{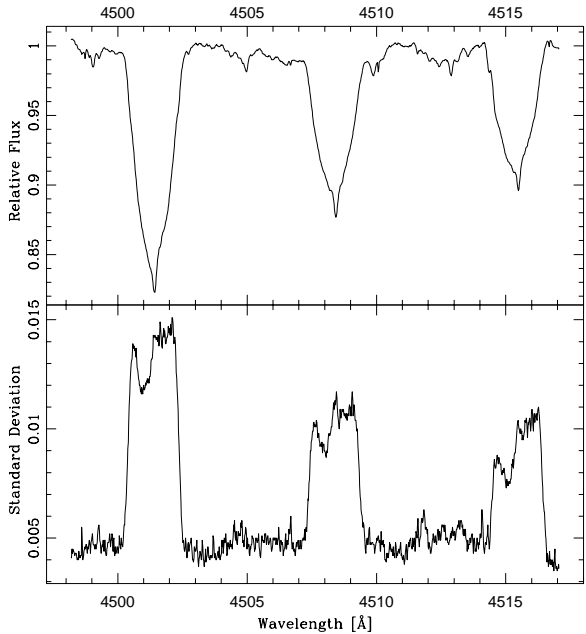}}
\caption[]{Upper panel: average spectrum normalized to the continuum intensity.
Lower panel: standard deviation of individual spectrograms about the average 
one} 
\end{figure}

The analysis of the spectroscopic variations has been made on the two lines
TiII 4501 and FeII 4508, which are the best defined in our spectra and
free of strong blends. The two equilibrium equivalent widths are 15.8 and 11.9
 \kms~ respectively. In the following, when averages of quantities derived from
the two lines are computed, a weight proportional to their equivalent widths 
has been assigned.

A non--linear least squares fit of a rotationally broadened gaussian
profile to the average profiles of the two lines, allowed us to estimate
the projected rotational velocity and the intrinsic
widths:  $v\sin i= 69.0\pm1.0$ \kms~ and $W_{4501}=11.9\pm 1.0 $ \kms~ and 
$W_{4508}=15.4\pm1.0$ \kms.
This $v\sin i$ estimate is in excellent agreement with the value
of $70\pm 2$ \kms~ derived  in Paper II.
\section{Analysis of photometric data}
The frequency analysis of the photometric data was performed by means of the
least--squares technique (Vani\^cek 1971, see also Mantegazza 1997). 
The data in the $V$ and the $B$
band were first analyzed separately and yielded a solution with 16 and 15
terms respectively.

In order to improve the signal--to--noise ratio, $V$ and $B$ data were 
put together
by shifting and rescaling the $V$ time series following a procedure
similar to that described by Breger et al. (1998).
The resulting file with 1805 data points has been frequency analyzed in the
way previously described. 17 terms were singled out and are listed in the
last column of table 1.
\begin{figure*}[]
\resizebox{17cm}{!}{\includegraphics{8984.f2}}
\caption[]{The upper part of each panel shows the differential V 
lightcurves with respect to the star's mean
brightness. The brightness increases upward. 

The lower part of each panel shows the radial velocity variations (\kms).

Open circles: observations; solid line: synthesized curves derived from the 
best--fitting models.}
\end{figure*}

The modes are listed in order of detection and the $S/N$ ratios were computed
using the Period98 package (Sperl, 1998). According to Breger et al.
(1995) peaks with $S/N>4$ have a high probability of being intrinsic to 
the star.
We see that this
criterion is met by 15 out of 17 detected periods.
For some of them there is the uncertainty of 1 \cds in particular for
$f_{11}$, $f_{14}$, $f_{17}$. The solution reported here is the combination 
of values which
give the least--squares best fit.
In the same table we report for comparison the corresponding frequency
values detected by our previous campaigns.
The comparison between the frequencies in common in different seasons
tells us that their values agrees within 0.005 \cds
(in the worst cases their difference is of about 0.01 \cd).
For $f_{11}$ and $f_{17}$ this comparison  confirms the
values here assumed, while for $f_{14}$ the closest connection is with
an 1 \cds alias of 7.17 \cds term. However the distance between
the two terms (0.06 \cd) is an order of magnitude larger than the above
reported average separation between all the other terms, so this connection
is very doubtful.
All the other terms are in agreement, with the exception of $f_{5}$, whose old
value was 7.667 \cd. Since the amplitude in the more recent data is larger
and the analysis of the spectroscopic data (see below)
favours the choice of the 6.66 \cd, in the following we shall adopt this
as the true frequency.

\begin{table}
\caption{Modes detected from the lightcurve analysis, their $B$ and $V$ 
amplitudes,
and comparison with the detections of the 1992 campaign}
\begin{tabular}{c c c c c c c }
\hline
                     & Freq.  &  S/N    &  A(B)  & A(V)  &  Old Freq. \\
                      &  [c/d] &         &  [mmag]& [mmag]&   [c/d] \\
\hline
$f_1$                &  7.394 &  128   &  50.8 &  37.2&     7.393 \\
$f_2$                &  6.036 &   25   &   9.5 &  7.6 &    6.042  \\
$f_3$                & 13.988 &   24   &   9.2 &  7.1 &   13.983  \\
$f_4$                &  7.453 &   19   &   7.5 &  5.6 &    7.465  \\
$f_5$                &  6.656 &   13   &   5.1 &  3.9 &    7.667  \\
$f_6(\simeq f_2+f_4)$ & 13.517 &   13   &   5.1 &  4.0 &   13.517  \\
$f_7$                & 12.917 &   11   &   3.9 &  3.3 &  (14.913?)\\
$f_8 = f_1+f_2$      & 13.434 &    8.8 &   3.2 &  2.8 &   13.437  \\
$f_9$                & 11.771 &    7.0 &   2.9 &  1.9 &   11.767  \\
$f_{10} = 2f_1$      & 14.779 &    7.0 &   2.9 &  2.0 &   14.779  \\
$f_{11}$             &  5.882 &    5.8 &   2.1 &  1.8 &    5.886  \\
$f_{12} = 2f_2$      & 12.092 &    5.5 &   1.9 &  1.8 &   12.088  \\
$f_{13}$             & 10.622 &    4.5 &   1.9 &  1.2 &   ---     \\
$f_{14}$             &  6.221 &    4.0 &   1.4 &  1.4 &   (7.160?)\\
$f_{15}$             &  8.848 &    4.3 &   1.6 &  1.4 &   ---     \\
$f_{16}$             &  9.003 &    3.5 &   2.0 &  0.6 &   ---     \\
$f_{17}$             & 11.264 &    3.4 &   1.2 &  0.9 &   11.272  \\
\hline
\end{tabular}
\end{table}

About the two weakest terms, which are only marginally significant,
we have to note that they have been independently detected from the
analysis of line profile variations (see below), and furthermore
the 11.26 \cds term was independently detected in the
1992 photometric data: this considerably strengthens their
reliability. Regarding the 9.003 \cds term, its spectroscopic counterpart was
detected at 10.01 \cds. Since the diagram describing the changing of the phase
of this perturbation across the line profiles looks cleaner by considering
10.01 instead of 9.003 \cds, we adopted the former value as the correct one.

Fig. 2 shows in the upper part of each panel the differential V measurements
with respect to the star's mean
brightness, and the best fitting 17--sinusoid model: no significant 
systematic differences between fitting curve and data points are apparent.
 
The availability of three V datasets allows us to compare
the evolution in time of the mode amplitudes.
In Table 2 we report the amplitudes in the three seasons of the 9
strongest modes.
It is apparent that $f_3$, $f_5$, $f_6$ and $f_7$ have shown conspicuous 
variations, while $f_1$, $f_2$ and $f_4$ seem substantially stable.

\begin{table}
\caption{$V$--light amplitudes of the strongest photometric modes in
three different epochs. The variations of $f_3$,$f_5$, $f_6$ and $f_7$ are remarkable}
\begin{tabular}{c c c c c}
\hline
  & Freq. & \multicolumn{3}{c}{V amplitudes}\\
  & [\cds] & \multicolumn{3}{c}{[mmag]}\\
   &       & 1989 & 1992 & 1996 \\
\hline
$f_1$ &  7.39 & 36.6 & 36.1 & 37.2 \\
$f_2$ &  6.04 &  7.7 &  6.7 &  7.6 \\
$f_3$ & 13.99 &  3.2 &  3.5 &  7.1 \\
$f_4$ &  7.46 &  5.2 &  5.0 &  5.6 \\
$f_5$ &  6.66 &  1.9 &  1.3 &  3.9 \\
$f_6$ & 13.52 &  0.7 &  4.8 &  4.0 \\
$f_7$ & 12.92 &  1.8 &  1.3 &  3.3 \\
$f_8$ & 13.43 &  3.2 &  2.4 &  2.8 \\
$f_9$ & 11.77 &  1.2 &  2.5 &  1.9 \\
\hline
\end{tabular}
\end{table}

\section{Analysis of line profile variations}

\subsection{Radial velocities}

Radial velocities were derived by the computation of the line barycenters
(this is equivalent to evaluate the first moment of the line profile).
They have been derived for both lines and then a weighted mean has been
performed. The resulting time series has been Fourier analyzed by means
of the same least--squares technique adopted for the light curve analysis.
13 frequency terms have been detected, all corresponding to analogous terms
found from the analysis of photometric data or to their 1 \cds aliases and one
more term with a frequency close  to 1 \cds, which was also detected in the
subsequent analysis of the line profile variations (see below).
Table 3 reports in successive columns, arranged in order of increasing
frequency: the adopted frequency, the frequency detected in the radial
velocity power spectrum, the amplitude of the term, the order of detection.
\begin{table}
\caption{Modes detected from radial velocity analysis}
\begin{tabular}{cccc}
\hline
\multicolumn{2}{c}{Frequency}&&\\
 Adopted & Observed & Amplitude & Order of\\
\multicolumn{2}{c}{[\cds]}& [\kms]& detection\\
\hline
   -- & 0.96 & 0.3 &  8\\
 5.89 & 4.92 & 0.2 & 13\\
 6.04 & 6.04 & 0.7 &  3\\
 6.22 & 5.18 & 0.2 & 10\\
 6.66 & 6.63 & 0.5 &  4\\
 7.39 & 7.39 & 3.9 &  1\\
 7.47 & 7.47 & 0.5 &  6\\
10.01 & 8.00 & 0.2 &  9\\
11.26 &10.23 & 0.1 & 14\\
11.77 &11.76 & 0.2 & 11\\
12.09 &11.15 & 0.2 & 12\\
13.44 &13.49 & 0.4 &  7\\
13.52 &13.56 & 0.5 &  5\\
13.98 &13.97 & 0.9 &  2\\
\hline
\end{tabular}
\end{table}

The observed radial velocities are shown in the lower part of each panel of 
Fig. 2 together their best fitting curve. In this case too no systematic
differences between the data and the computed curves are apparent.

\subsection{Mode detection}

The search for periodicities in the line profile variations has been performed
by applying the least squares power spectrum technique generalized to study
line profile variations. A detailed description of the technique can be
found in Mantegazza and Poretti (1999).
Fig. 3 reports in the top panels the power spectra of the line profile
variations in the TiII 4501 line (left panel) and in the FeII 4508 line (right
panel) and in the bottom ones the residual power spectra, obtained
giving as ``known constituents'' (defined in Mantegazza \& Poretti, 1999)
all the detected terms. Possibly some more
modes are still present at frequencies below 15 \cds, but at present they 
are not detectable without ambiguity.
\begin{figure*}[t]
\resizebox{17cm}{!}{\includegraphics{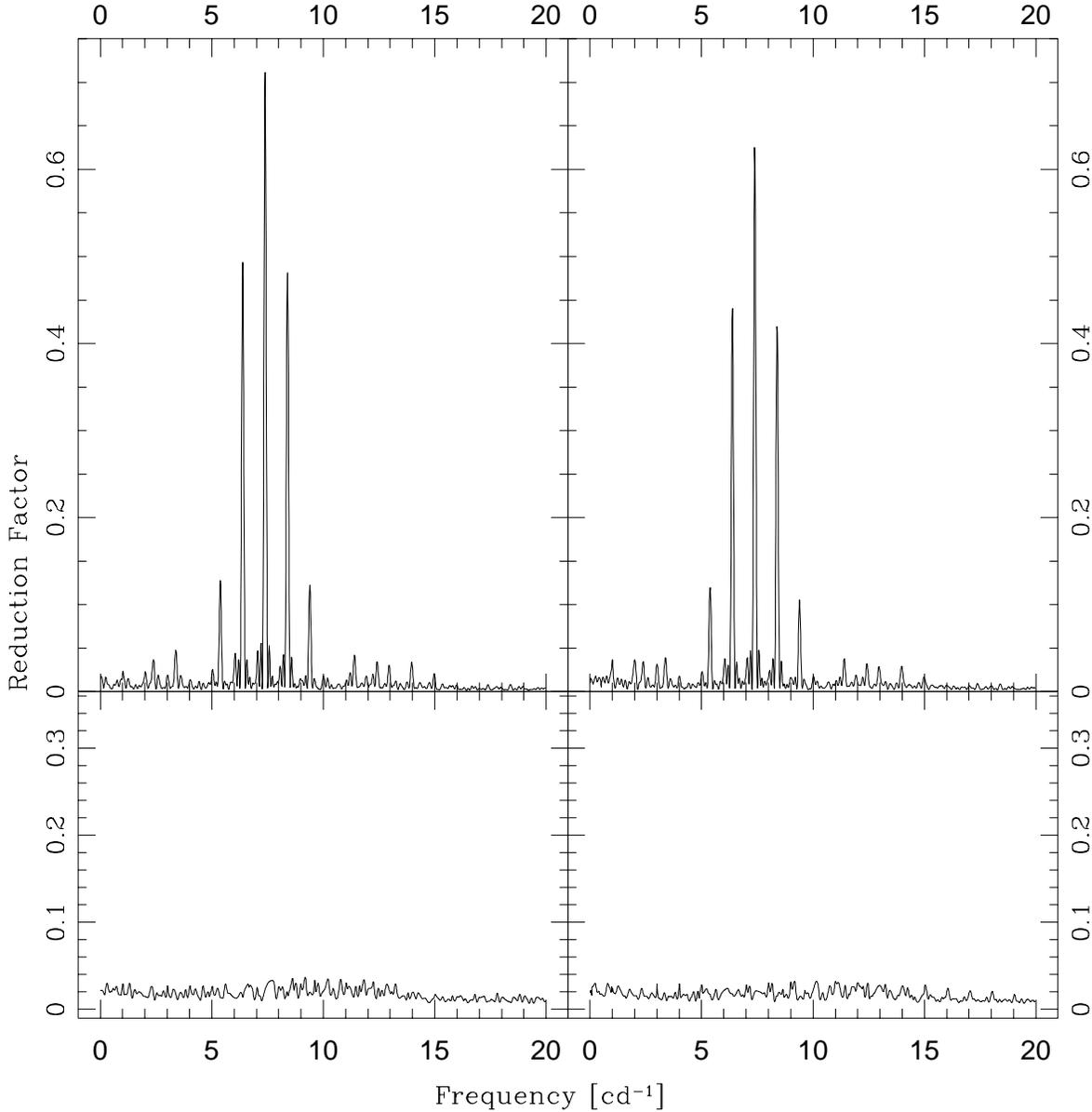}}
\caption[]{Least--squares power spectra of the two lines (left TiII 4501, right
FeII 4508).
Top panels: original power spectra; bottom  panels:
spectra with all the detected modes given as ``known constituents''.}
\end{figure*}

Table 4 reports the detected terms in order of detection for the TiII 4501
line (which has the stronger $S/N$). In successive columns we report:
the derived frequency in the TiII 4501 line, that in the FeII 4508 line,
the frequency of the corresponding photometric term, the finally
adopted frequency for the mode, and the rms amplitude along the line
profile of the spectroscopic term. 
This amplitude has been derived from a weighted average between the
amplitudes of the two lines normalized to the intensities of the TiII line
 (the rms amplitudes in the two lines are about in the ratio 1.8:1) and it
is measured in per mille of the continuum intensity.
\begin{table}
\caption{Frequencies derived by line profile variation analysis compared 
with the
photometric ones and the final adopted values for the successive mode 
identification}
\begin{tabular}{c c c c c }
\hline
\multicolumn{4}{c}{Frequency}& Spect.\\
TiII & FeII  &Phot.   & Adopted  &Ampl. \\
\multicolumn{4}{c}{[\cds]} & $ $ \\
\hline
 7.39& 7.39& 7.39& 7.39&14.9\\
 6.04& 7.04& 6.04& 6.04& 2.7\\
13.97&12.95&13.98&13.98& 3.0\\
 2.01& 2.00& --  & --  & -- \\
12.24&12.24&11.26&11.26& 1.8\\
 9.99& 9.02& 9.01&10.01& 2.4\\
 ---  & 0.56&--&   --  & -- \\
12.91&12.91&12.92&12.92& 2.1\\
 6.63& 6.63& 6.66& 6.66& 1.7\\
 5.89& 5.91& 5.88& 5.89& 1.6\\
14.78&14.78&14.78&14.78& 1.6\\
12.09& --  &12.09&12.09& 1.7\\
10.64& 9.65&10.62&10.62& 1.5\\
13.54&13.54&13.52&13.52& 1.7\\
12.45& --  &13.44&13.44& 1.4\\
\hline
\end{tabular}
\end{table}

We see that the three strongest terms are the same found from photometry,
namely 7.39, 13.98, and 6.04 \cd.
There is also a low frequency term (2.0 c/d) which is an alias of the
reciprocal of the sidereal day and probably it is due to an instrumental
effect (a similar term has been found and discussed
by Mantegazza and Poretti, 1999, analysing the spectrograms
of BB Phe). The analysis of the line profile variations of the FeII 4508 line
shows also a low frequency term at 0.56 \cds, which however has not been
detected in the other higher $S/N$ line. It is therefore not very likely that
this signal can be attributed to the star.

%

We can see that all the spectroscopically detected terms have
been also photometrically detected. This tells us that in X Cae 
we don't have to expect
the presence of high--degree modes  as found
in other $\delta$ Scuti stars (see e.g. the cases of BV Cir, Mantegazza
et al. in preparation, HD 101158 (Mantegazza 1997) and $\tau$ Peg, 
Kennelly et al. 1998).

An indepenent check of these results has been performed using the CLEAN
algorithm  generalized to the analysis of the line profile variations.
Examples of the use of this technique to analyse line profile variations
 in $\delta$ Scuti
stars can be found in De Mey et al.(1998) and Bossi et al.(1998).

\begin{figure*}[t]
\resizebox{17cm}{!}{\includegraphics{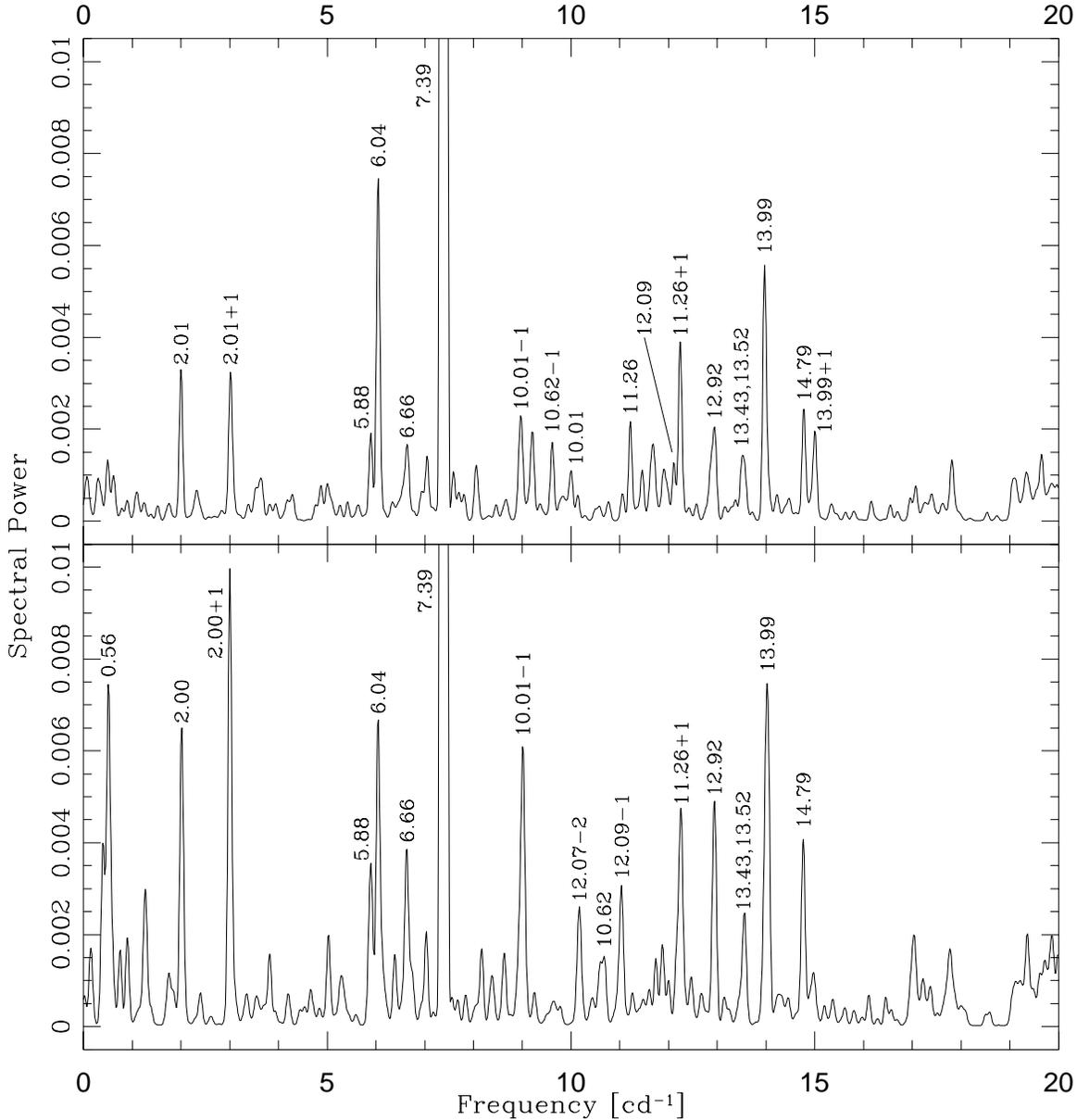}}
\caption[]{CLEAN power spectra of the two lines. TiII 4501 (top), and FeII 
4508 (bottom).
The labels indicate the peaks corresponding to the terms detected
with the least-squares analysis or to their 1 \cds aliases. The 7.39 \cds peak
is off scale.}
\end{figure*}

The original spectra have been computed
up to 30 c/d, but for clarity the figures are truncated to 20 c/d, since there
is no significant signal above this value, 
and have been obtained with a gain of 0.3 and
500 iterations.
Figure 4 shows the spectra of the two lines with the frequency
identification of the highest peaks which correspond to those detected by the
least--squares analysis or to their  1 \cds aliases.
Finally Fig. 5 shows a comparison between the relative amplitudes of the 
modes detected from photometry 
(top panel), radial velocity variations (middle panel) and
line profile variations (bottom panel).

\begin{figure}[t]
\resizebox{9cm}{!}{\includegraphics{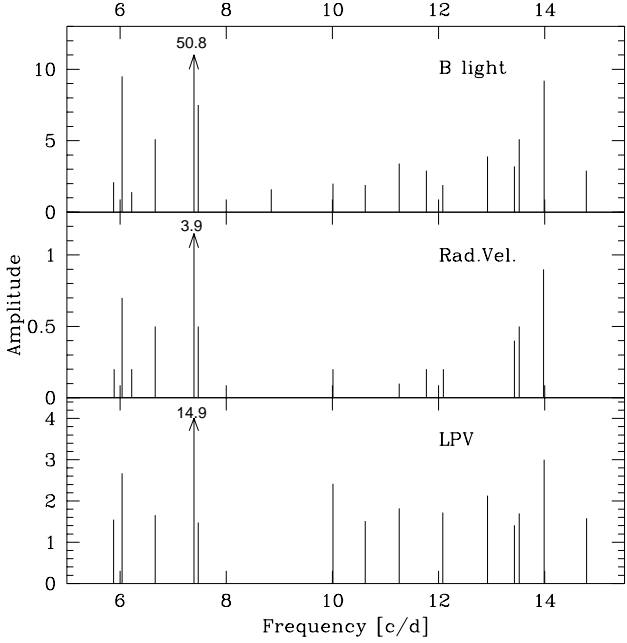}}
\caption[]{Comparison between the amplitudes of the modes
detected from lightcurve (upper panel, mmag), radial velocity variations
(middle panel, \kms) and
variations along the line profile (lower panel, rms amplitude averaged
on the two lines and normalized to the intensity of the TiII line,
and expressed in per mille of the continuum intensity).
The amplitudes of the dominant mode (7.39 \cds)
are off scale, therefore they are marked with an arrow and given as
labels in each panel.}
\end{figure}

\subsection{Amplitude and phase diagrams}

After the detection of all the relevant modes affecting the line profiles,
it is possible to obtain their amplitudes and phases across the line profile
($A_j(\lambda)$,
$\phi_j(\lambda)$ $j=1,\# modes$ with their formal errors
$\delta A_j(\lambda)$ and $\delta\phi_j(\lambda)$)
by means of a simultaneous least--squares fit ( Mantegazza \& Poretti, 1999). 
We also get the
estimate of the unperturbed average line profile (the zeropoints of
each pixel time series fit: $A_0(\lambda)$).

\begin{figure*}[t]
\resizebox{17cm}{!}{\includegraphics{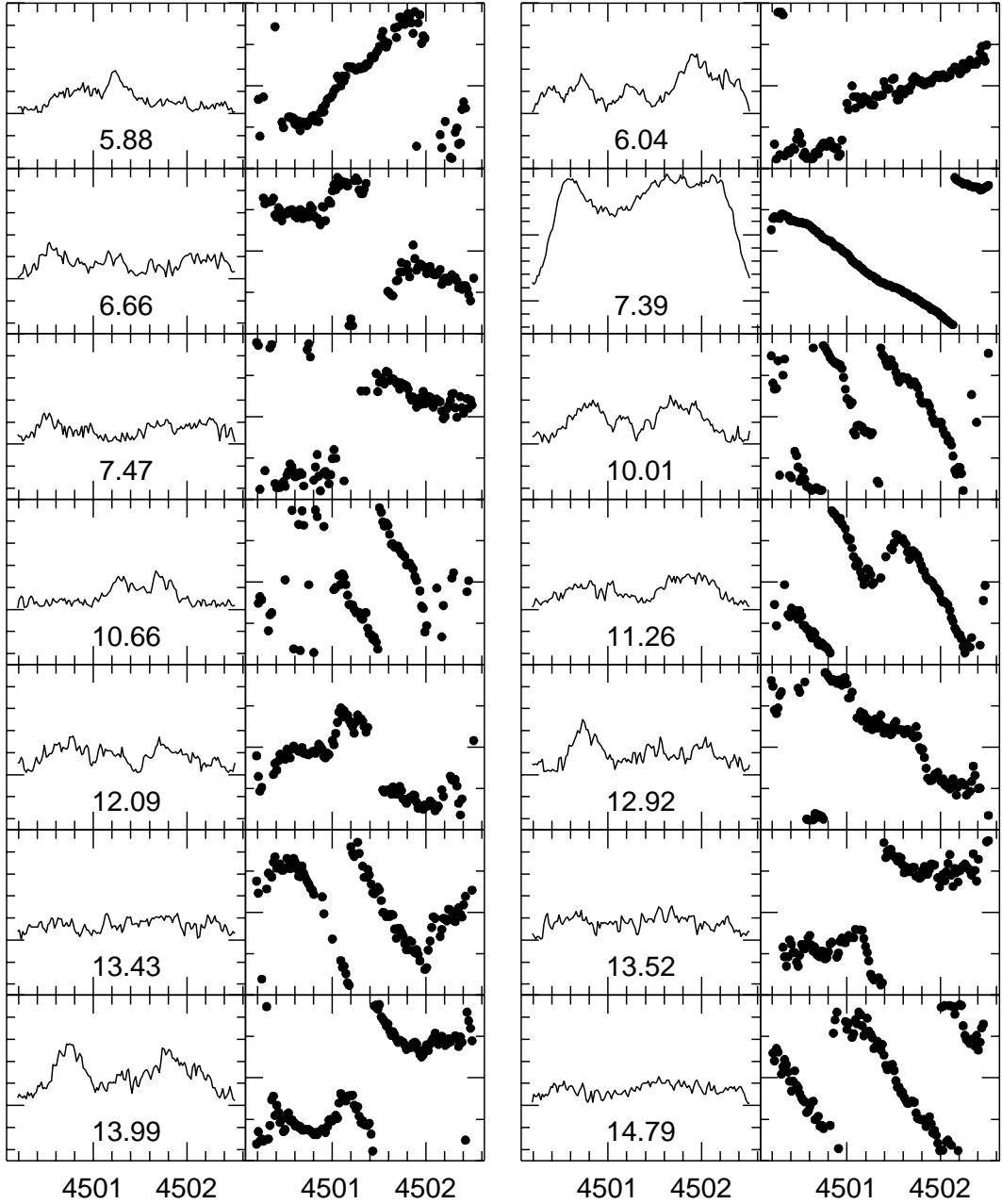}}
\caption[]{Amplitudes and phases across the line profile of the TiII 4501 line
of all the detected modes.
The ticks on the amplitude diagrams correspond to 0.002 units in the continuum
intensity. Please note that the scale of the 7.39 \cds term is different from
the others.
For the phase diagrams each tick corresponds to $50^o$. }
\end{figure*}
Figure 6 shows the behaviour of amplitude and phase of each mode
across the profile of the TiII 4501 line.
To obtain this figure we used all the terms detected in the spectra
 with the addition of the 7.47 c/d one. This is among those photometrically
strongest, and it has been also detected in the radial velocity data.
Probably its spectroscopic detection is hampered by its closeness
to the dominant mode, so that it is drowned in its peak because 
the ratio between the amplitudes of the two modes is more unfavourable than
in the light and radial velocity curves.

Very similar results have been obtained from the fit of the FeII 4508 line.
These diagrams, in particular the phase ones,
give already some clues about some possible mode
identifications. We note that a mixture of different pulsation modes is 
present.
 The phase curve of the 5.88 \cds term is typical
of a low degree retrograde mode, while those of 7.47, 13.52 and 13.99 \cds 
indicate axisymmetric modes and that of  7.39 \cds indicates a
low--degree prograde mode.

\subsection{Model fit to line profile variations}

It is possible to try to estimate the $\ell,m$ parameters of each detected
mode by fitting the variations it induces on the line profile shape
(Bossi et al., 1998; Mantegazza \& Poretti, 1999).

In order to do this it is necessary to separate the contributions of the
different modes. The perturbations $\Delta p_j(\lambda,t)$
induced by mode $j$ on the line profile 
can be approximated as:
$$\Delta p_j(\lambda,t)=A_j(\lambda)\cos(2\pi\nu_jt+\phi_j(\lambda))
\eqno (1)$$
with the corresponding error $\delta\Delta p_j(\lambda,t)$ derived from error
propagation by $\delta A_j(\lambda)$ and $\delta\phi_j(\lambda)$.
These last quantities as well as  $A_j(\lambda)$ and $\phi_j(\lambda)$
have been derived in the previous subsection.
We can try to fit these functions ($\Delta p_j(\lambda,t)$) with perturbations 
computed with a model
of a non-radial pulsating star viewed at a certain inclination $i$.
So for each plausible choice of $\ell,m,i$ we can build a discriminant
$$\sigma_p(\ell,m,i) = \sum_{\lambda}\sum_t{(\Delta p_j(\lambda,t)-
\Delta p_c(\lambda,t,\ell,m,i))^2\over{\delta\Delta p_j(\lambda,t)^2}} 
\eqno(2)$$
where $\Delta p_c(\lambda,t,\ell,m,i)$ are the computed profile variations
which best fit the observed ones.

If moreover we have simultaneous photometric observations, we can obtain,
by simultaneously fitting all the terms detected in the light curve,
 amplitude and phase with respective errors for the $j$ mode. Therefore
we can calculate its light variations and relative errors
($l_j(t)$ and $\delta\l_j(t)$) and compare them
with those predicted by the best fitting models and obtain the discriminant:
 $$\sigma_l(\ell,m,i) =\sum_t(l_j(t)-
l_c(t,\ell,m,i))^2/\delta\l_j(t)^2 \eqno(3)$$

A global discriminant is then defined as:
$$ \sigma_T(\ell,m,i)= \sigma_p(\ell,m,i)+\sigma_l(\ell,m,i) \eqno(4)$$

This is the function minimized with a non--linear least--squares
fit for each detected $j$ mode and for any choice of $\ell,m,i$.

In summary, since by means of eq. 1 we estimated the variations induced on the
line profile by each individual mode, we can compare them with the
variations computed from a model of a pulsating star for each 
possible non--radial mode. Such a
comparison is done computing the weighted sum of the squares of
the differences between the two quantities: we called such a sum
the {\em discriminant}. This approach is similar to the one
developed to identify the pulsation modes with the ``moment
method'' (Balona, 1986, 1987; Aerts, 1996).

In addition, since the model allows us to compute
the light variations related with each mode, we can then adjust its
free parameters to best fit at the same time observed line profile
and light variations. This can be done with a non--linear
least--squares algorithm that minimizes the {\em global
discriminant} described in eq. 4.

The process of deriving the line profile and light variations for
each mode (eq.1) carries on an error estimation together with the
results. The squared reciprocal of such estimated errors are the
weights to be used in the discriminants computation, with the
additional advantage to give to the dicriminants themselves the
adimensional character needed to add them in eq. 4.

The model used to compute the synthetic line profile variations
($\Delta p_c(\lambda,t,\ell,m,i)$) and light variations ($l_c(t,\ell,m,i)$)
is the 
one described by Balona (1987). For each assigned {$\ell,m,i$} the
computed profile variations can be modeled according to the amplitude and
phase of vertical ($v_r$) and horizontal ($v_r$) velocities and flux 
variations.
For $\delta$ Scuti stars usually $v_h \ll v_r$ and therefore in order to not
introduce into the model too many free parameters, we keep the usual
theoretical relation (e.g. Heynderickx et al. 1994)
$v_h=74.4Q^2 v_r$ ($Q$ pulsation constant) and $\psi_h=\psi_r$.

The observed light variations constrain strongly the computed flux variations,
so it is very useful to have simultaneous spectroscopic and photometric data,
otherwise in order to get meaningful physical results it is better to
fit a simplified model which considers velocity variations only, neglecting
flux variations.

We applied the method to the data by computing eq. 1 for each identified 
mode.
In order to do this ten equi-spaced phases which cover the corresponding 
pulsation period were used.

We have explored for each detected term all
the possible modes with $0\le\ell\le4$, and considered inclinations between 
20 and 90 degrees.
The independent analysis of the two spectral lines gives concordant indications
about the better fitting modes, and moreover the estimated physical parameters
of these modes are similar.
Fig. 7 shows the discriminants of the best fitting modes for all the
detected frequency terms as a function of the inclination of the rotational
axis. In order to allow a comparison between the different panels the
discriminant scale is the same for all of them, but it
has been shifted in order that the best fitting mode discriminant in 
each panel was close to the lower border. 
The discriminants are the weighted means of those derived
from the fits of line profile variations in the two spectral lines.
The figure contains  also the panels for the terms at 12.09, 13.43, 13.52  and
14.79 \cds  which, as already remarked, could not be independent
pulsation modes. 
\begin{figure*}[t]
\resizebox{17cm}{!}{\includegraphics{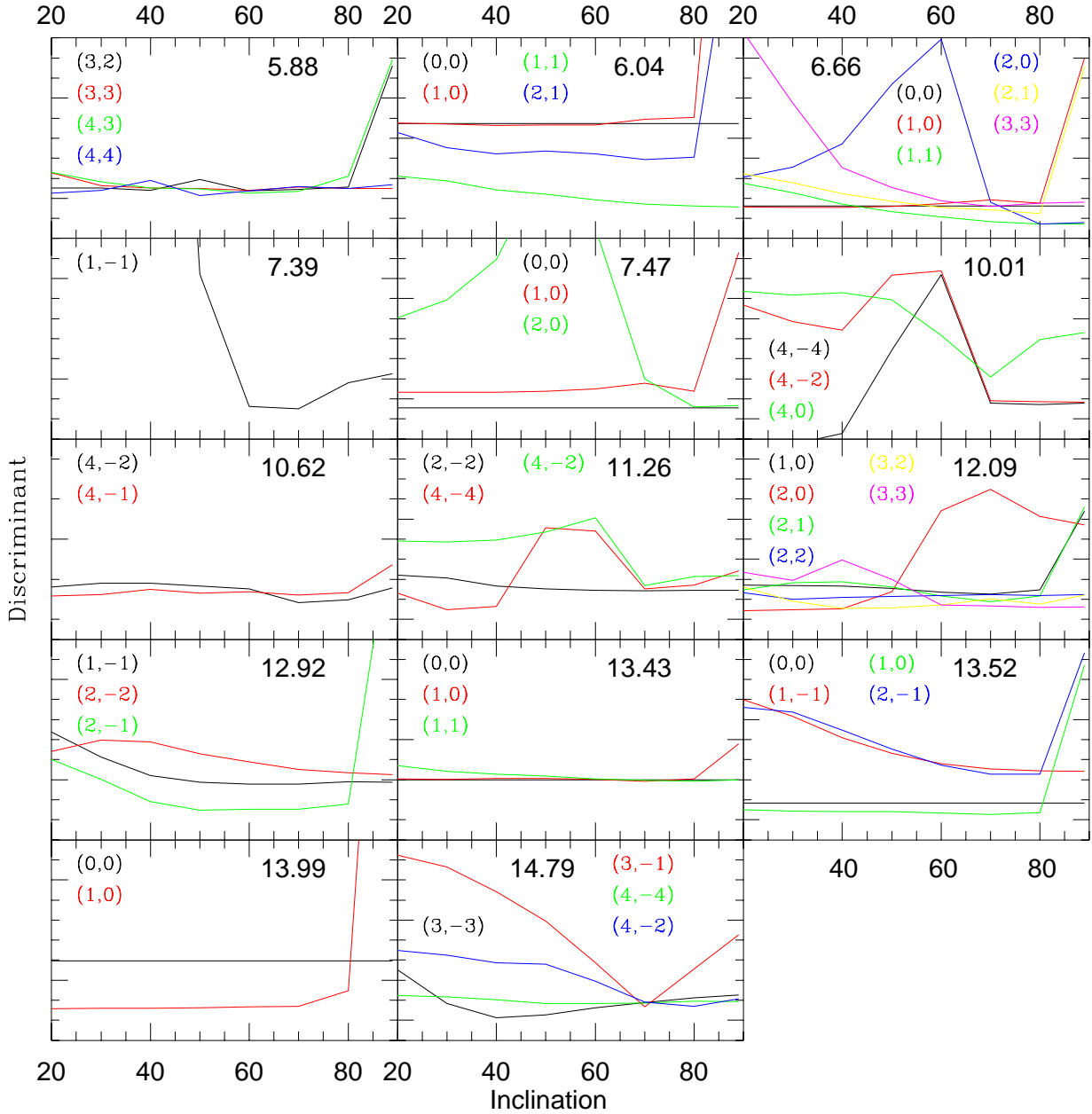}}
\caption[]{The discriminant of the best fitting modes are shown for each
detected frequency term as a function of the inclination of the rotational
axis. $\ell,m$ numbers are in the same colour of the corresponding line.}
\end{figure*}

In order to evaluate the inclination, a total discriminant has been
computed by adding for each inclination the minimum global discriminants 
of all the
detected modes. As a matter of fact, the 7.39 \cds mode dominates this
discriminant and all the other modes supply only perturbations.
Their presence affects appreciably this discriminant only for $i$ close
to $90^o$.
The result is reported in Fig. 8, where we see
that there is a rather broad minimum centred at about $70^o$.
\begin{figure}[t]
\resizebox{9cm}{!}{\includegraphics{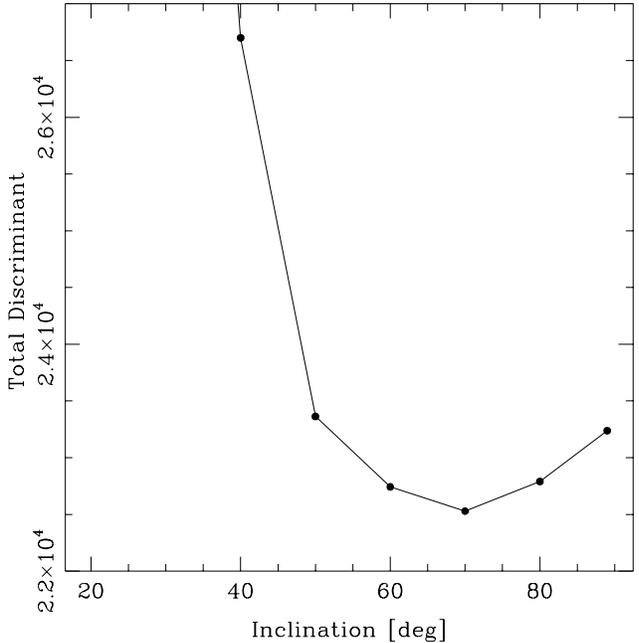}}
\caption[]{Total discriminant obtained summing the best discriminants
of all the modes at each inclination versus the inclination angle.}
\end{figure}

Assuming this inclination, 
the equatorial rotational velocity results of about 73 \kms, and therefore
the rotational period is about 2.4 d and the ratio between rotational and
pulsation frequencies is lower than about 0.06 for all the modes.

Having established a reasonable value for the inclination angle, we can
look at the discriminants of the individual modes in order to decide
which are the more plausible identifications for this inclination.
The proposed $\ell,m$ identifications for each detected
mode are reported in the $6^{th}$ column of Table 7.
For some modes these identifications are rather uncertain or there are 
different solutions which fit almost equivalently well the line profile
variations.
The more uncertain results are given between brackets.

In Fig. 9 we show the variations of the Ti 4501 line profile due to
the 7.39 \cds term
(solid line) phased on a complete pulsation cycle
and the corresponding best fitting variations generated by a model of a
$\ell=1,m=-1$ mode. This model explains the 93\% of the line profile
variance, fits the $B$ light variations with a standard deviation
of 0.4 mmag and the radial velocity variations with an accuracy of $\pm$0.2
\kms.
\begin{figure}[t]
\resizebox{9cm}{!}{\includegraphics{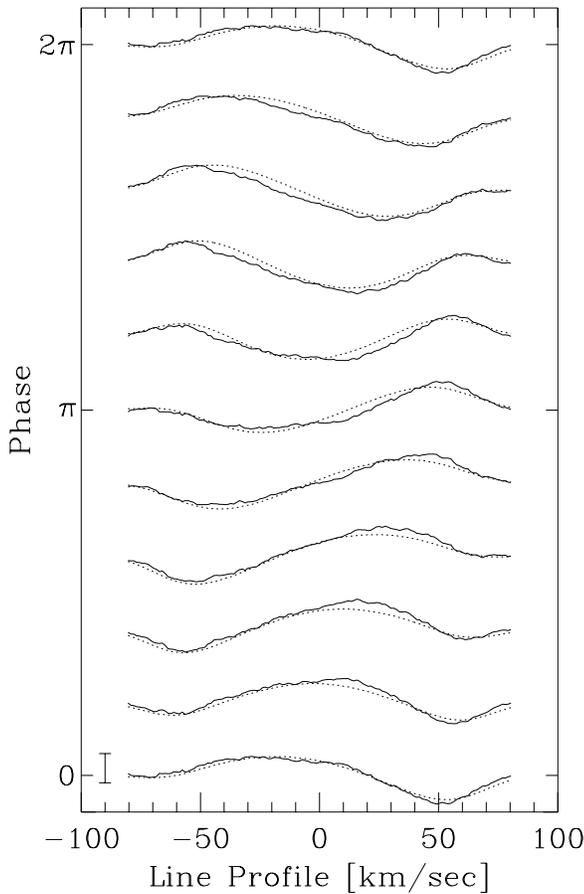}}
\caption[]{The variations induced on the profile of the TiII 4501 line
by the 7.39 \cds term phased over 1 cycle (solid lines) and the best fitting
model of a mode with $\ell=1, m=-1$ and $i=70^o$ (dashed lines).
The small bar at the lower left indicates a 0.02 amplitude in continuum
intensity units.}
\end{figure}

In Fig. 10 we show the observed amplitude (upper panel) and phase
 variations (bottom panel) of the 7.39 \cds term
across the TiII line profile with the error bars derived from the
least--squares fit, together with the corresponding curves derived 
from the best fitting model (solid lines).
If we take into account all the approximations present in our model
of pulsating star, we have to be more than satisfied by the quality of
these results. Furthermore the figure makes it clear that the bump in the
amplitude curve at the left
of the line profile, that was already present in the 1992 data (see
Fig. 4 of paper II), is a signature of the pulsation mode.
\begin{figure}[t]
\resizebox{9cm}{!}{\includegraphics{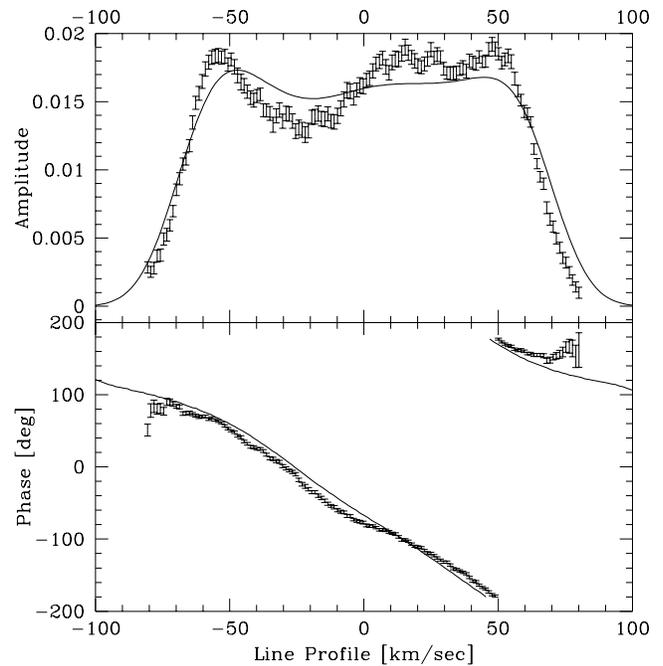}}
\caption[]{Observed (bars) and computed (solid lines) behaviours across
the TiII 4501 line profile
of the amplitude (top panel) and phase (bottom panel) of the 7.39 \cds mode.
The computed model
correspond to the best fitting $\ell=1,m=-1$ mode, $i=70^o$.}
\end{figure}
\subsection{Color phase shifts}

In order to add weight to the mode identification a further independent
tessera can be included in the mosaic by considering the phase
differences between $V$ and $B$ light variations.
Figure 11 shows this quantity versus the respective observed frequency
for the 6 strongest photometric modes 
(namely, in order of frequency, 6.04, 6.66,
7.49, 7.47, 13.52 and 13.58 \cd). 
From the line profile variation analysis we have derived that apart from 
the 7.47 \cds mode, all the others
have as preferred choice $\ell=1$ (see Table 5). 
We see from the diagram that all of them
fall in a strip of $\pm 3^o$ about the zero shift (dotted lines), just as
expected.
On the contary the 7.47 \cds mode, for which the best option is $\ell=0$,
has the highest fase difference and falls just outside this strip, again as
expected.
\begin{figure}[t]
\resizebox{9cm}{!}{\includegraphics{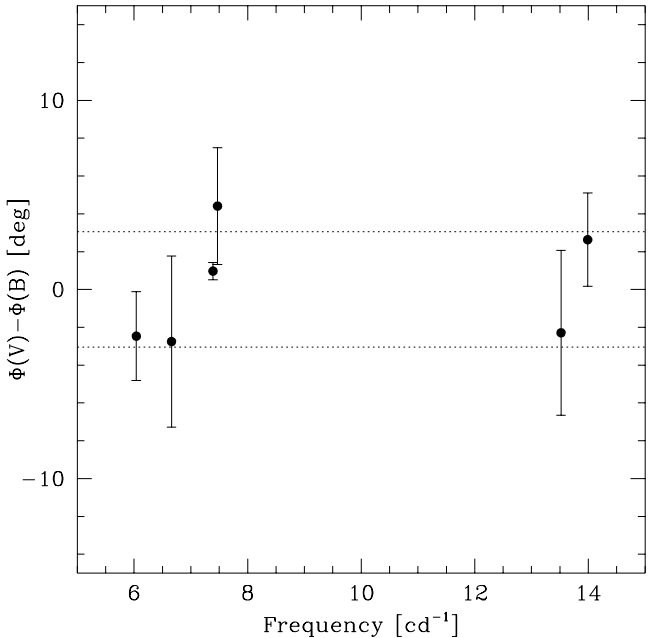}}
\caption[]{Phase shifts between $V$ and $B$ light variations for the 6
strongest photometric modes. The dashed lines define the strip in which
falls all the modes with $\ell=1$.}
\end{figure}

\section {Discussion}
\subsection{Physical quantities related to pulsation modes}

In the Table 5 we present the physical quantities obtained by
fitting line profile and light variations for the better identified modes. 
In the successive columns
we give: observed frequency ($\nu$), $\ell,m$,
frequency in the stellar reference frame
($\nu_0$),
pulsation constant ($Q$), amplitude of local radial velocity ($v_r$),
percentual local flux variation ($\Delta F/F$) at the line wavelength, 
phase shift between flux and radial variation ($\psi$), percentual local radius
variation ($\Delta R/R$), and the ratio between relative flux and radius 
variation ($f$).
The reported amplitudes are sinusoidal amplitudes, so that the full amplitudes
are twice these values.
\begin{table*}
\caption{Physical parameters of the better identified modes}
\begin{tabular}{r l r c r r r c r }
\hline
$\nu$ & $\ell,m$&$\nu_0$ &$Q$& $v_r$&$\Delta F/F$&$\Psi$&$\Delta R/R$& $f$\\

       [\cds]&      & [\cds]& [d]&[\kms] & \% & [deg] & \% &\\
\hline
 5.88 & 3,2  &  6.71 & 0.033 &  2.2 &  3.9 &  96 & 0.20 & 20 \\
 6.04 & 1,1  &  6.46 & 0.035 &  2.9 &  4.0 & 121 & 0.25 & 16 \\
 6.66 & 1,1  &  7.09 & 0.032 &  2.0 &  1.8 & 119 & 0.17 & 11 \\
 7.39 & 1,-1 &  6.98 & 0.032 & 25.1 & 19.0 & 114 & 2.04 &  9 \\
 7.47 & 0,0  &  7.47 & 0.029 &  2.4 &  2.0 & 102 & 0.18 & 11 \\
11.26 & 2,-2 & 10.44 & 0.021 &  1.7 &  1.4 & 132 & 0.09 & 16 \\
12.92 & 2,-1 & 12.51 & 0.018 &  6.3 &  5.1 & 145 & 0.29 & 18 \\
13.52 & 1,0  & 13.52 & 0.016 &  5.1 &  3.6 & 149 & 0.22 & 16 \\
13.99 & 1,0  & 13.99 & 0.016 &  9.5 &  7.2 & 164 & 0.39 & 18 \\
\hline
\end{tabular}
\end{table*}

The mode with the best defined parameters is obviously 7.39 \cd. We believe
that these parameters are rather reliable: the formal uncertainties that we 
got from the
simultaneous fit of its line profile and light variations
are $\pm 1$~\kms~ on $v_r$, $\pm1.7\%$ on $\Delta F/F$, $\pm7^o$
on $\Psi$, $\pm0.09\%$ on $\Delta R/R$ and $\pm 1.0$ on $f$.
It is interesting to compare the $f=9$ and $\Psi=114^o$
values with those predicted
by theoretical models for a star with physical parameters similar to X Caeli.
For low--degree $\ell$ modes of low radial order we have, for
a mixing length parameter $\alpha=0.5$, $f\sim11$ and $\Psi\sim115^o$ and, for
$\alpha=1.0$, $f\sim8$ and $\Psi\sim150^o$
(theses data were kindly supplied us by W. Dziembowsky, and the $f$ values 
have beeen rescaled at the line wavelength).
The agreement with the theory is 
rather good, and it seems that our data favour a mixing length parameter 
of about 0.5.

The parameters of the other modes are more uncertain, given the small
amplitudes of the perturbations produced by these modes on the line profiles.

It is worth noticing that the two lowest frequency modes (5.89, 6.04 \cds)
are retrograde, as probably is also the case for the 6.66 \cd~one .
Up to now there was not a large evidence of the
presence of retrograde modes in $\delta$ Scuti stars: Kennelly et al. (1998)
suspected that two high--degree modes in $\tau$ Peg could be retrograde,
and Mantegazza and Poretti (1999) showed the presence of a rather certain
low--degree retrograde mode in BB Phe.

We observe too, that the 6.66 and 7.39 \cds terms have, within
the approximations, due to the uncertaintinies on $i$ and on the Ledoux
factor $C$,
the same frequency in the stellar reference frame. Hence, having
$\ell=1$, and $m=1$ and $m=-1$ respectively, they could be the rotational
splitting of the same mode.

\subsection{Fundamental stellar parameters as derived from pulsation}

The identification of the 7.47 \cds mode allows us to estimate 
some physical parameters of the star. As we have seen, this mode
has been clearly detected in the light and in the radial velocity
curves, but its detection is more uncertain in the line profile variations. 
Moreover its phase curve
shows a clear rotation of $180^o$ in the line center, and the amplitude of the
line profile variations are larger in the wings. These facts suggest 
that we are in presence
of a mode that shifts the line rather than affecting its shape.
Since in addition the mode is preferably seen in the integrated
quantities and the inclination of the rotational axis is rather high,  
this mode is probably a radial mode.
The same suggestion is supplied by its colour phase shift, even if on this
basis alone we cannot rule out the $\ell=1$ possibility (Fig. 11).

If this is the correct identification
then its $Q$ value as derived from the assumed physical
parameters ($0.029\pm0.004$ d) would put it midway between fundamental 
($Q=0.033\pm0.001$ d)
and first overtone mode ($Q=0.025\pm0.001$ d). In the first hypothesis the
7.39 \cds term should be a $\ell=1$ $f$ mode, while according to the second it
should be a 
$\ell=1$ $p_1$ mode (see the theoretical models by Fitch, 1981).
In Table 6 we report the stellar fundamental parameters and their uncertainties
 as derived from three different hypotheses:
a) adopting the Hipparcos parallax, b) 7.47  \cds is the radial fundamental 
mode, c) 7.47  \cds is the $1^{st}$ overtone radial mode. 
All the hypotheses assume
the effective temperature derived from multicolor photometry, and the validity
of the theoretical models by Shaller et al. (1992).
In this table $\nu_F$ is the fundamental radial mode frequency. The uncertainty
on the 7.47 \cds mode ($\pm$0.02 \cd) is larger than on the other detected 
modes (see section 4) because of its proximity to the 7.39 \cds mode.

\begin{table}
\caption{Physical parameters of the star estimated by considering the 7.47 \cds
mode as the radial fundamental or the first overtone mode and compared with
the parameters derived using the Hipparcos parallax}
\begin{tabular}{c c c c}
\hline
          & Hipparcos &Fund. & 1st ov.\\
\hline
$T_{eff}$   & $6900\pm 100$& $6900\pm 100$ & $6900\pm 100$ \\
$M/M_{\odot}$& $2.0\pm0.1$  & $1.90\pm0.1 $ & $1.95\pm0.05$ \\
$M_{bol}$   & $1.30\pm0.15$ & $1.47\pm0.08$ & $1.06\pm0.08$ \\
$R/R_{\odot}$& $3.45\pm0.32$ & $3.16\pm0.08$ & $3.82\pm0.10$ \\
$\log g$    & $3.66\pm0.09 $& $3.72\pm0.04$ & $3.57\pm0.04$ \\
$\rho/\rho_{\odot}$& $0.049\pm0.014$&$0.061\pm0.004$&$0.036\pm0.002$\\
$\nu_F$[\cds]   & $6.7\pm0.7$  & $7.47\pm0.02$ & $5.75\pm0.02$\\
d[pc]    &$102\pm7$&$93\pm3$ &$113\pm4$\\
\hline
\end{tabular}
\end{table}

The bolometric magnitude obtained with the fundamental mode assumption is
 $1\sigma$ higher than derived from Hipparcos parallax, while that obtained
with the
first overtone assumption is $1.4 \sigma$ lower. On these bases it is difficult
to get a firm conclusion even if it is slightly favoured the fundamental mode
option for the 7.47 \cds term. This without considering the possible presence
of a circumstellar envelope that, introducing some extinction, could 
shift the result toward the first overtone mode hypothesis.
In any case it can be appreciated how the use of the astroseismological 
information leads to a shrinking of the error bars.

\begin{table*}
\caption{Complete list of the frequency terms detected in X Cae }
\begin{tabular}{r r r c r l c c}
\hline
freq. & \multicolumn{4}{c}{Amplitude}&$\ell,m$&  Alt.Freq. & Possible\\
      &  V  & B-V & RV & LP  &   && couplings \\

[\cds] & \multicolumn{2}{c}{[mmag]}& [\kms]&[$10^{-3}$]&[\cds]\\
\hline
5.89  &  1.8  &0.3 &0.2   &1.6  &  3,2       &         &\\
6.04  &  7.6  &1.9 & 0.7  & 2.7 &  1,1       &         &\\
6.22  &  1.4  &0.0 & 0.1  & --- &  ---       &   7.16  &\\
6.66  &  3.9  &1.2 & 0.4  & 1.7 &  1,1; 2,1  &  (7.67) &\\
7.39  & 37.2  &13.6&  3.9 & 14.9&  1,-1      &         &\\
7.47  &  5.6  &1.9 & 0.5  & 1.5 &  0,0       &         &\\
8.85  &  1.4  &0.2 & ---  & --- &  ---       &         &\\
10.01 &  0.6  &1.4 & 0.2  & 2.4 & (4,-4;4,-2)&   9.01  &\\
10.62 &  1.2  &0.7 & ---  & 1.5 & (4,-2;4,-1)&         &\\
11.26 &  0.9  &0.3 & 0.1  & 1.8 &  2,-2      &  12.26? &\\
11.77 &  1.9  &1.0 & 0.2 &  --- &  ---       &         &\\
12.09 &  1.8  &0.1 & 0.2&   1.7 &  2,1; 1,0  &         &    ($\approx 6.04*2$)\\
12.92 &  3.3  &0.6 & --- &  2.1 &  2,-1      &         &\\
13.44 &  2.8  &0.5 & 0.4 &  1.4 &  ---       &         &    6.04+7.39\\
13.52 &  4.0  &1.1 & 0.5 &  1.7 & 1,0; 0,0   &         &  ($\approx 6.04+7.47$)\\
13.99 &  7.1  &2.1 & 0.9 &  3.0 &  1,0       &         &\\
14.78 &  2.0  &0.9 & --- &  1.6 &  ---       &         &    7.39*2\\
\hline
\end{tabular}
\end{table*}

\section{Conclusions}

The analysis of the simultaneous photometric and spectroscopic observations
of X Caeli has supplied the following results:
\begin{itemize}
\item{} 17 terms have been detected in the photometric data, most of which
were already detected in the 1989 and 1992 campaigns. The comparison of
the amplitudes of the strongest terms has shown that while the dominant mode
(7.39 \cds) looks rather stable other terms have large amplitude variations.
\item{}The analysis of the radial velocity curve detected 13 terms,
and again 13 (but not exactly the same)
were detected from the analyis of the line profile variations.
All these terms are among those detected from photometry.
Therefore no high--degree modes have been detected, even if the line profiles
are sufficiently broadened by rotation to allow it.
\item{}By means of simultaneous least--squares fits of the line profile and 
light variations induced by each mode we were able to supply $\ell,m$ 
identifications for many modes,
and to estimate the inclination of the rotational axis (about $70^o$).
\item{}The two shortest period modes are retrograde ones, while the
6.66 and 7.39 \cds modes are probably the result of the rotational splitting
of a $\ell=1$ mode.
\item{}The physical parameters of the
 modes with the most reliable identifications have been given (Table 5). 
Rather accurate values have been 
obtained for the dominant mode, which has $\ell=1$, $m=-1$. When compared
 with those predicted by theoretical models they
allow the estimation of the mixing length parameter which results about
0.5.
\item{}The 7.47 \cds mode is probably radial. It remains uncertain if it is
the fundamental or the first overtone. The stellar physical parameters
derived in both hypotheses are given and compared and discussed with those
obtained using the Hipparcos parallax.
\end{itemize}

Table 7 summarizes all the detected modes in the present campaign with the
different techniques and in order of increasing frequency.
In the successive columns 
the $V$, $B-V$, RV amplitudes and the rms one across the line profile 
are given.
The last three columns give furthermore the most probable $\ell,m$
identification, the eventual alternative frequency due to aliasing, and
the possible relationships with the observed frequencies of other modes.

The research presented in this paper has supplied a lot of
useful and interesting results concerning the pulsational
behaviour and the physical characteristics of the $\delta$ Scuti
star X Caeli. However a number of questions remains open and needs
further investigations. First of all some of the detected terms
still suffer of the 1 \cds uncertainty which could probably be
removed through a multisite photometric campaign.

Furthermore a better resolution both in spectroscopy and
photometry would be useful to have a better plot of some close
terms as the 7.39 and 7.47 \cds. This would allow us to get a more
accurate evaluation of the contribution of each mode to the
observed variations and consequently would allow us to perform a more
reliable mode identification. For instance the confirmation of the
radial nature of the 7.47 \cds mode would be of particular
relevance in order to nail the fundamental stellar physical
parameters.

Longer baselines would moreover allow us to detect more pulsation
modes whose presence can be guessed from the present data, and to
improve the accuracy of the phase shifts in different photometric
bands and hence to allow an independent check of the $\ell$
identifications supplied by spectroscopy.

Given the brigthness of the object the use of a larger telescope
in photometry  would not significantly improve the results.
However, since some modes produce variations of only a few
thousands of the continuum intensity (see Fig. 6), a larger
telescope in spectroscopy would allow us to obtain more accurate
estimates of the line profile variations through higher $S/N$ data.

Nonetheless we emphasize that in order to improve the mode
detection an effort should be made on the theoretical side too.
Indeed, as Fig. 10 clearly shows, the computed amplitudes across
the line profile of the best fitting mode ($1,-1$) for the
dominant term (7.39 \cd), for which the spectroscopic $S/N$ is amply
adequate, have still considerable systematic deviations from the
observed amplitude variations.

\begin{acknowledgements}
We are grateful to Dr. W. Dziembowsky for supplying us the theoretical data,
to Dr. M. Bossi for some enlightening discussions, to Dr. E. Poretti for
a critical reading of the manuscript and to Dr. V. Fraschini for the help in
the data reduction.
Finally we wish to thank the referee, Dr. D. Sullivan, for his useful 
comments and suggestions. 

\end{acknowledgements}

\end{document}